# Using A Negative Spatial Auto-correlation Index to Evaluate and Improve Intrinsic Tag Map's Multi-Scale Visualization Capabilities

Zhiwei WEI [a,b] & Nai YANG [c]

[a]*Guangdong Laboratory of Artificial Intelligence and Digital Economy, Shenzhen, China*; [b]*The Aerospace Information Research Institute, Chinese Academy of Sciences, Beijing, China*; [c]*School of Geography and Information Engineering, China University of Geosciences, Wuhan, China*.

Address for correspondence: Zhiwei Wei. E-mail: 2011301130108@whu.edu.cn

The two authors contributed equally to this research.

# Using A Negative Spatial Auto-correlation Index to Evaluate and Improve Intrinsic Tag Map's Multi-Scale Visualization Capabilities

**Abstract**: The popularity of tag clouds has sparked significant interest in the geographic research community, leading to the development of map-based adaptations known as intrinsic tag maps. However, existing methodologies for tag maps primarily focus on tag layout at specific scales, which may result in large empty areas or close proximity between tags when navigating across multiple scales. This issue arises because initial tag layouts may not ensure an even distribution of tags with varying sizes across the region. To address this problem, we incorporate the negative spatial auto-correlation index into tag maps to assess the uniformity of tag size distribution. Subsequently, we integrate this index into a TIN-based intrinsic tag map layout approach to enhance its ability to support multi-scale visualization. This enhancement involves iteratively filtering out candidate tags and selecting optimal tags that meet the defined index criteria. Experimental findings from two representative areas (the USA and Italy) demonstrate the efficacy of our approach in enhancing multi-scale visualization capabilities, albeit with trade-offs in compactness and time efficiency. Specifically, when retaining the same number of tags in the layout, our approach achieves higher compactness but requires more time. Conversely, when reducing the number of tags in the layout, our approach exhibits reduced time requirements but lower compactness. Furthermore, we discuss the effectiveness of various applied strategies aligned with existing approaches to generate diverse intrinsic tag maps tailored to user preferences. Additional details and resources can be found on our project website: https://github.com/TrentonWei/Multi-scale-TagMap.git.

## 1. Introduction

With its visually appealing appearance and straightforward creation process, the tag cloud has gained widespread popularity beyond analytical contexts (Viegas et al., 2009). Numerous software applications and websites have emerged to facilitate the generation of tag clouds, such as Wordle (Feinberg, 2009), ManiWordle (Koh et al., 2010), Word Cloud Explorer (Heimerl et al., 2014), EdWordle (Wang et al., 2018), emordle (Xie et al., 2023) and WordArt[①]. This surge in popularity has piqued the interest of the geographic research community, leading to frequent utilization of tag clouds for visualizing opinions or activities associated with geographic locations or areas, commonly referred to as tag maps (Yang et al., 2019). Among these tag maps, intrinsic tag maps are the most prevalent, wherein tags are fitted within geographic boundaries (Yang et al., 2023). In this study, our focus is specifically on this type of map.

Although intrinsic tag maps can be considered specialized versions of tag clouds, standard tag cloud layout algorithms are unsuitable for these polygon-based tag maps due to the irregular geographic boundaries. Consequently, several algorithms have been proposed to address the layout creation for intrinsic tag maps. For instance, Nguyen and Schumann (2010) introduced the Taggram technique, which arranges size-varied and colorized tags alphabetically along the main vertical skeleton of the geographical region. However, the tag orientations are fixed in their approach. More recently, Buchin et al. (2016) proposed a greedy algorithm for tag placement, organizing tags orderly according to their sizes with various orientations possible. Martin and Schuurman (2017) utilized the wordle technique to generate intrinsic tag maps with tags placed along a helix. Meanwhile, Yang et al. (2019) introduced a triangulated irregular network (TIN) to subdivide geographical regions into triangle subareas, with each triangle's centroid serving as a potential location for tag placement, allowing for orientation adjustments based on the triangles.

Although the aforementioned algorithms can generate tag layouts with tags in fixed or unfixed orientations, they primarily address tag layouts at a specific scale. However, conventional exploration often entails users navigating across multiple scales, leading to potential dissatisfaction with tag layout consistency across different scales (Zhang et al., 2018). For instance, intrinsic tag maps typically adopt a multi-scale visualization approach by dynamically adjusting tag sizes and removing smaller tags, which can result in uneven distributions of the remaining tags. Figure 1 illustrates this issue. Figure 1(a) depicts an example intrinsic tag map generated at level 1 for Ethiopia using the web-based application Tag Map Explorer developed by Yang et al. (2019). In Figures 1(b) and 1(c), which represent lower levels of the tag map, large empty areas (Areas A, C, and D) or close proximity between large tags (Area B) may occur if smaller tags at level 1 are removed. This discrepancy arises from the uneven placement of tags with varying sizes within the region. Therefore, to facilitate multi-scale intrinsic tag map visualization, it is imperative to devise a tag layout that ensures an even distribution of tags of sizes across the region at different scales.

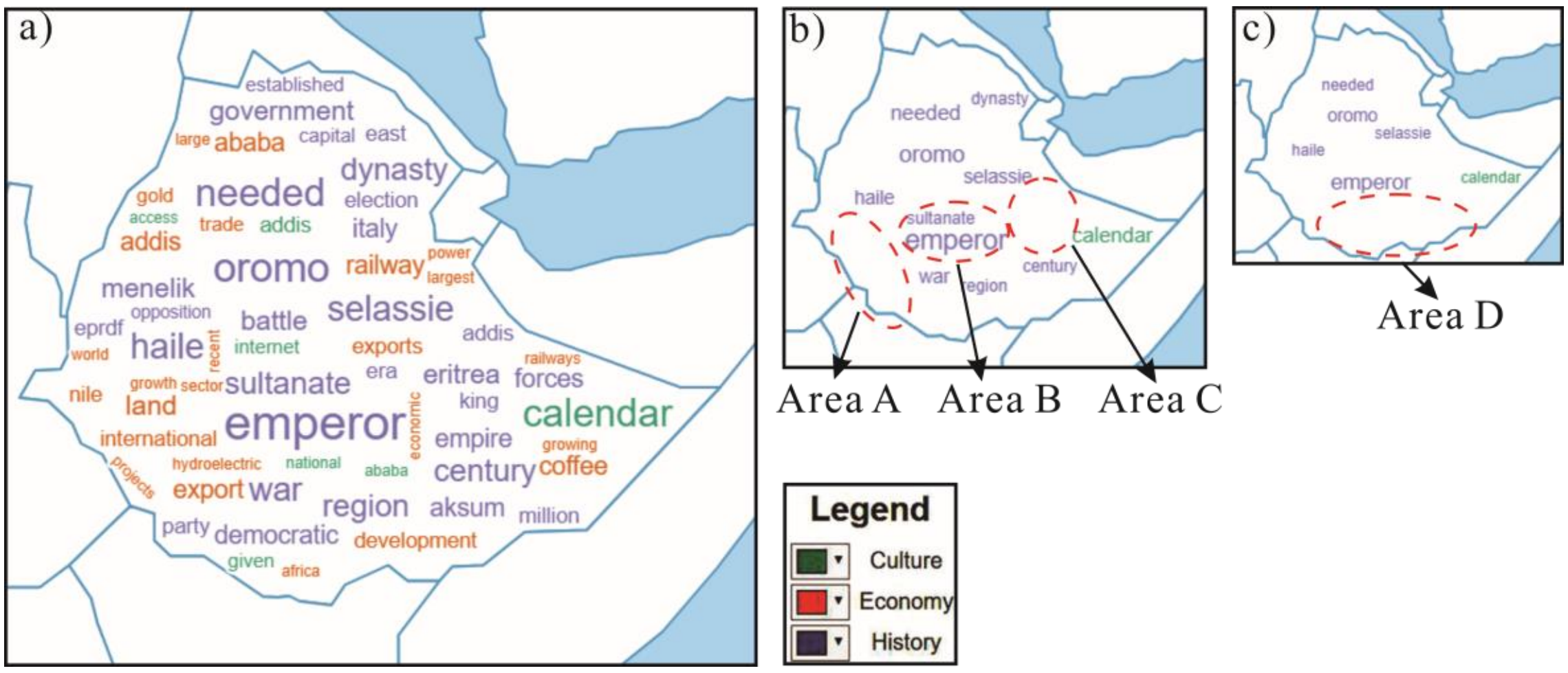

Figure 1. Multi-scale intrinsic tag maps which were generated by Tag Map Explorer (Yang et al., 2019). a) Level 1; b) Level 2; c) Level 3.

To generate a tag layout with an appropriate distribution of tag sizes for multi-scale visualization, our initial step involves developing an index to assess the distribution. Spatial autocorrelation indices are commonly employed to gauge the geographic distribution of data values (Getis and Ord, 1992). Accordingly, we can leverage a spatial autocorrelation index to evaluate the distribution of tag sizes. As highlighted by Wang et al. (2019) in the context of map generalization, a distribution where map features of varying sizes are evenly scattered across a region can ensure that larger features maintain a relatively even distribution after smaller ones are removed. Similarly, as smaller tags are dynamically removed from the tag map layout when users navigate to smaller scales, ensuring that tags of different sizes are evenly scattered in the layout can also help larger tags maintain a relatively even distribution after the removal of smaller tags. This distribution phenomenon is commonly referred to as negative spatial autocorrelation (Griffith and Aibia, 2019). Therefore, instead of using indices that focus on positive spatial autocorrelation, we adopt a negative spatial autocorrelation approach to assess the distribution of tag sizes. Subsequently, we can employ this index to enhance the tag layout for multi-scale visualization while employing an existing algorithm to generate the layout of an intrinsic tag map.

Motivated by the above thoughts, we try to apply the negative spatial auto-correlation index to evaluate the tag size distribution in an intrinsic tag map layout. As analyzed above, an optimal tag size distribution indicated by this index can maintain uniform tag sizes. This prevents large empty areas and avoids close proximity between large tags after the removal of small tags when users navigate to smaller scales, thereby enhancing the multi-scale visualization capability. Subsequently, we integrate this index into an existing approach, namely, the TIN-based tag map layout method proposed by Yang et al. (2019), to augment its support for multi-scale visualization. Unlike their method, which iteratively identifies suitable locations for each remaining tag by only considering the area of triangular subareas, we select the optimal location for a tag in each iteration by considering both the area of the triangular subareas and the capability for multi-scale visualization. Consequently, our method yields a more refined tag map layout conducive to multi-scale visualization.

## 2. Related works

### 2.1 Intrinsic tag map visualization

The origins of tag map-like visualizations can be dated back to 1741 when Gottfried Hensel visualized the world's languages and alphabets in Synopsis Universae Philologiae (Wikimedia Commons, 2012). Another milestone was Milgram and Jodelet's (1976) mental map of Paris 1976, often cited as the earliest instance of tag maps. However, these early tag maps were manually drawn. With the emergence of Web 2.0 and advancements in computer science, many automated systems and algorithms were developed to generate diverse tag maps. These maps can be broadly classified into overlay tag maps, extrinsic tag maps, intrinsic tag maps, and variant tag maps (Yang et al., 2023). Intrinsic tag maps, fitting a tag cloud inside a geographic boundary to emphasize association with specific administrative regions, are the primary focus of this study.

Nguyen and Schumann (2010) introduced the Taggram technique, which arranges tags alphabetically along the primary vertical skeleton of the geographical region. Building on this foundation, Nguyen et al. (2011) further refined the Taggram approach, developing several strategies for spatial-temporal visualization. These strategies encode temporal information by altering the visual appearance of text or incorporating additional visual artifacts associated with the tags. Similarly, Tag@Map, developed by De Chiara et al. (2012a, 2012b), constructs a map containing a tag cloud representing summarized data extracted from the underlying dataset. This tool facilitates expert user interaction to analyze phenomena of interest. However, both Taggram and Tag@Map maintain fixed tag orientations in their approaches. In contrast, Buchin et al. (2016) proposed a greedy algorithm for tag placement, prioritizing graphic design over data depiction. Tags are organized orderly based on their sizes, with orientations adaptable to various configurations. Their method attempts to utilize tag arrangements to depict the spatial shapes of geographic regions without simultaneously visualizing the actual boundaries of these shapes, which may be inadequate for regions with complex geometries. In recent years, Martin and Schuurman (2017) utilized wordle.py to generate multi-scale intrinsic tag maps representing outputs from topic models applied to geolocated Twitter data. While effective, wordle.py typically positions tags along a helix, potentially leaving regions incompletely filled. Meanwhile, Yang et al. (2019) introduced a triangulated irregular network (TIN) approach to subdivide geographical regions into triangle subareas. Each triangle's centroid serves as a potential location for tag placement, enabling orientation adjustments based on the triangles' characteristics. Subsequently, Yang et al. (2020) further explored intrinsic tag maps, offering a comprehensive analysis of their utility and usability.

In a departure from conventional intrinsic tag maps, some researchers have also explored tag layouts by considering the spatial relevance of tags. Reckziegel et al. (2018) introduced a layout algorithm that aimed to position tags as close as possible to their related locations, enhancing the spatial relevance of the visualization. Similarly, Jiang et al. (2019) extended Reckziegel et al. (2018)'s approach to visualize the spatial distribution of ethnic populations. This thematic information focuses on the demographic composition of different locations, emphasizing the distribution and concentration of various ethnic groups within geographic regions. In their approach, the location of each tag is strategically determined to reflect the geographic location of the corresponding phenomenon within the regions.

In summary, while these algorithms produce satisfactory intrinsic tag maps for specific purposes, they primarily address tag layout at a specific scale. Navigating across multiple scales may lead to dissatisfaction with tag layout consistency (Zhang et al., 2018). Thus, it is necessary to evaluate and enhance existing algorithms to support multi-scale visualization.

### 2.2 Spatial autocorrelation indices

Spatial autocorrelation indices are widely used in spatial analysis to assess the degree of similarity between spatial observations and their neighboring locations (Griffith and Aibia, 2010). These indices help identify patterns of spatial dependence, clustering, or dispersion in geographic data. Various spatial autocorrelation measures exist, such as Global Moran's I Index (Moran, 1950), $G_i$* Statistic (Getis and Ord, 1992), and LISA (Anselin, 1995), each serving specific purposes in different contexts. Among these indices, the Global Moran's I Index evaluates overall spatial autocorrelation across an entire study area and is suitable for assessing the overall distribution of tag sizes.

The Global Moran's I Index compares the observed spatial pattern of a variable with what would be expected under spatial randomness. It accomplishes this by calculating the covariance between the values of the variable at different locations, taking into account both the values themselves and the spatial relationships between those locations. The key to calculating the Global Moran's I Index lies in capturing the spatial relationships and assigning weights to any two locations that have a defined spatial relationship. Therefore, we can also follow the calculation process to calculate the index for tag sizes in a tag map to evaluate its capability for supporting multi-scale visualization.

## 3. Evaluating multi-scale visualization capabilities of tag maps using spatial autocorrelation indices

Given that maintaining a uniform distance between tags is a fundamental requirement in an intrinsic tag map, our objective in this study is to utilize the Global

Moran's I Index to quantify the distribution of tag sizes to support multi-scale visualization in tag maps. The Global Moran's I Index serves as a metric to evaluate the distribution between spatial observations and their neighboring locations (Griffith and Aibia, 2010). The spatial observations in our approach correspond to the sizes of tags, while the locations represent the centers of these tags. To implement this methodology, we follow the calculation process outlined in Section 2.2, which entails defining (1) neighboring relations and (2) a weight matrix.

### 3.1 Define the neighboring relations

Irregular boundaries, often characterized by islands or holes, are prevalent in intrinsic tag maps (Yang et al., 2019). Consequently, tags within separate polygons must be treated distinctly. Our objective is to evenly distribute tags across several region polygons. Therefore, neighboring relations are defined as follows: tags within the same polygon are considered neighbors.

### 3.2 Define the weight matrix

Based on the defined neighboring relations, we subsequently determine the weight ($w_{ij}$) between tags $i$ and $j$ using the inverse distance metric, a commonly utilized weight definition in spatial autocorrelation index calculations (Gimond, 2019). Specifically, let $d_{ij}$ represents the distance between the centers of tags $i$ and $j$, their weight $w_{ij}$ is then defined as Equation (1). Note that $w_{ij}$ is not normalized in Equation (1) because our goal is to compute the spatial autocorrelation index, which requires only the normalization of the index itself, as defined in Equation (2).

$$w_{ij} = \begin{cases} 1/d_{ij} & \text{(neighbouring tags)} \\ 0 & \text{(non-neighbouring tags)} \end{cases} \tag{1}$$

### 3.3 Define the index

As the regions for the tag map may comprise multiple polygons, we initially compute a sub-index ($I_m$) for the tags within each polygon. Subsequently, we derive the final index ($I$) to gauge the capacity for multi-scale visualization through a weighted sum, taking into account the number of tags in each polygon. As discussed in Section 1, it has been demonstrated that stronger negative spatial autocorrelation of tag size distribution enhances the effectiveness of multi-scale visualization in tag maps. Therefore, to evaluate the capability of supporting multi-scale visualization of an intrinsic tag map, we incorporate the negative Global Moran's I Index. By multiplying the Global Moran's I Index by -1, as defined in Equation (3), a higher index value indicates a better capability for supporting multi-scale visualization. Suppose the regions for the tag map consist of $m$ polygons, with each polygon hosting $N_m$ tags, then the sub-index $I_i$ for the tags within a polygon, based on the aforementioned analysis, can be computed as Equation (2).

$$I_m = -\frac{N_m}{\sum\sum w_{ij}} \times \frac{\sum\sum w_{ij} * (x_i - \bar{x})(x_j - \bar{x})}{\sum (x_i - \bar{x})^2} \tag{2}$$

where $x_i$ represents the size of tag $i$, and $\bar{x} = \frac{\sum_{i=1}^{N_m} x_i}{N_m}$. The overall index ($I$) to assess the multi-scale visualization capability across all polygons is calculated as Equation (3). Since the index is calculated based on the centers of the tags, and considering that each tag occupies an area, perfect negative spatial autocorrelation cannot be achieved (Gimond, 2019). Consequently, the index typically yields low values. To mitigate this limitation, we enhance the index by applying a scaling weight of 10, which confines its range to [-10, 10].

$$I = \frac{\sum I_m * N_m}{\sum N_m} * 10 \tag{3}$$

## 4. Improving multi-scale visualization of tag maps using spatial autocorrelation indices

The existing algorithms typically position tags based on their weights or alphabetical order, utilizing heuristic approaches. Therefore, we also incorporate a heuristic constraint by integrating the capability to support multi-scale visualization into these algorithms. The TIN-based tag map algorithm introduced by Yang et al. (2019) offers user-friendly features and facilitates multi-scale visualization strategies. Thus, we extend their algorithm by integrating our proposed index to enhance its multi-scale visualization capability.

### 4.1 The principle of our approach

The TIN-based tag map approach proposed by Yang et al. (2019) uses an iterative method to place tags from the largest to the smallest. This method involves dividing geographical regions into triangular subareas, with the centroid of each triangle serving as a potential location for centering a tag, as illustrated in Figures 2(a) to 2(d). To determine the location of a tag in each iteration, Yang et al.'s (2019) approach primarily considers the area of triangles, evaluating potential tag locations from the largest to the smallest triangle. However, within each iteration, a tag may have several suitable candidate locations, and the triangle area alone may not suffice to make the optimal decision. For instance, when placing the tag 'algiers' in Figure 2, focusing solely on the area of the triangles might place 'algiers' near other large tags, as seen in Figure 2(d). In contrast, considering the capability for support of multi-scale visualization, particularly aiming to achieve better negative spatial autocorrelation of tag sizes (as defined in Equation 3), could potentially lead to more optimal placement of 'algiers', as depicted in Figure 2(e), where large tags are distributed more evenly. This placement reduces the risk of large tags becoming too close when users navigate to smaller scales, thereby enhancing multi-scale visualization. Therefore, we propose selecting the optimal location for each tag among all available candidates in each iteration, considering not only the area of the triangles but also the negative spatial autocorrelation of tag sizes.

**Our principle is as follows**: we extend Yang et al. (2019)'s algorithm by also subdividing geographical regions into triangular subareas to identify suitable locations iteratively for each remaining tag with maximum size. Unlike their method, we evaluate all suitable candidate locations for each tag at every iteration, selecting the best one by considering both their existing criteria and the negative spatial autocorrelation of tag sizes.

Given the iterative placement of tags in our algorithm, it is possible that there may not always be a sufficient number of tags for computing the predefined spatial autocorrelation index within the current tag layout. To overcome this limitation, we initially introduce virtual tags to populate the region, facilitating the calculation of the predefined spatial autocorrelation index (Section 4.2). These virtual tags are phased out during subsequent iterations. Moreover, not all locations identified in the TIN-based approach as suitable for placing the current tag are viable. Certain unsuitable candidates need to be removed to optimize the tag layout (Section 4.3). Finally, we select the best candidate by considering both the existing criteria and the negative spatial autocorrelation of tag sizes (Section 4.4).

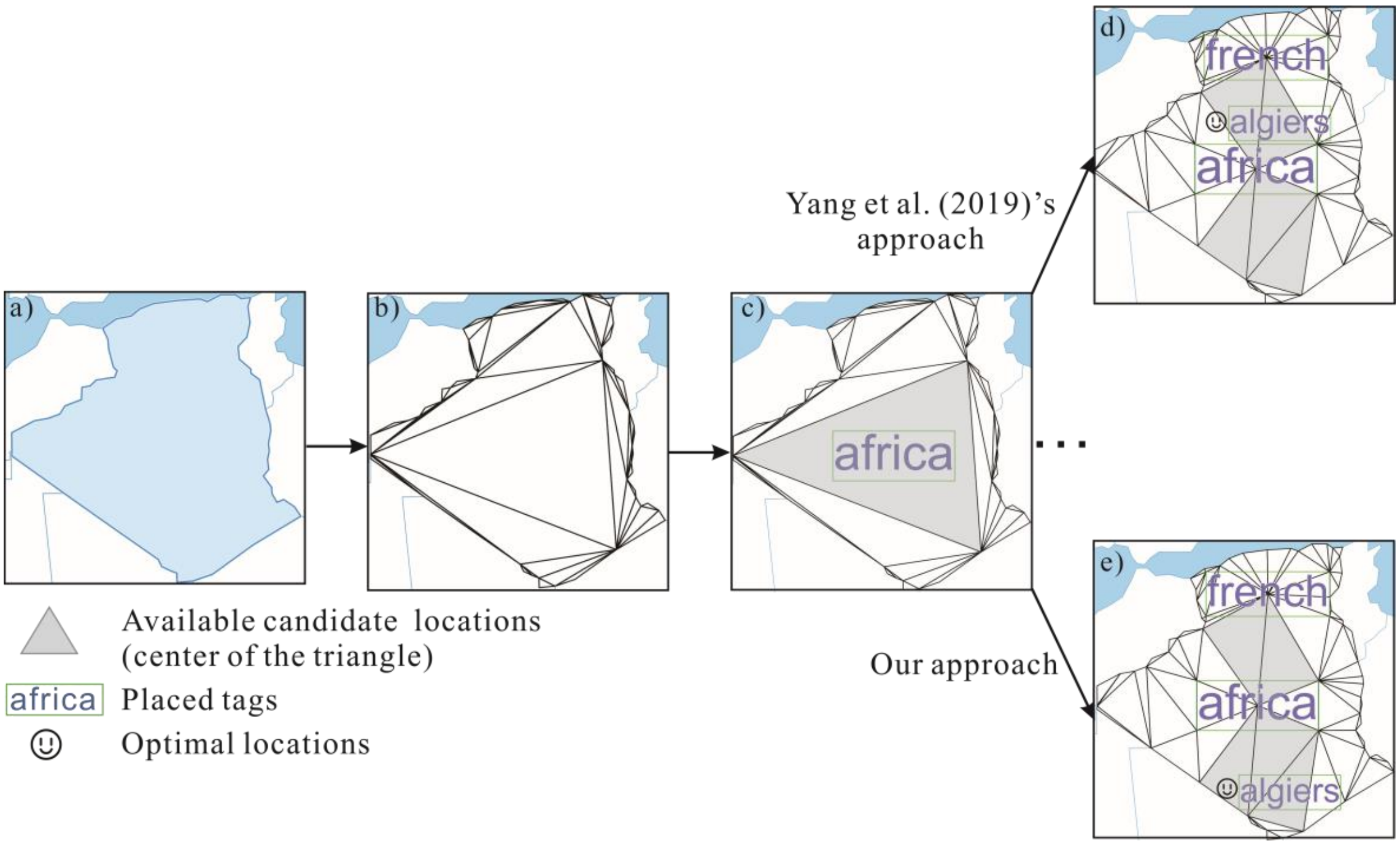

Figure 2. The principal of our proposed approach. (a) Region selected; (b) TIN constructed based upon the region outline; (c) Place the first tag; (d) Yang et al. (2019)'s approach: bounding box of the tags and their center points are added to reconstruct TIN, and place the tag within the triangular subarea with the largest area if available. (e) Our approach: Other locations within the triangular subareas would suffice, and we select an optimal location among all available candidates, taking into account both the area of triangles and the negative spatial autocorrelation of tag sizes.

### 4.2 Create virtual tags

Virtual tags are introduced to occupy the region polygons, ensuring a consistent spatial distribution for computing the predefined spatial autocorrelation index. When no tags are initially placed, the region polygons can be considered homogeneous. Consequently, the generated virtual tags should also maintain this homogeneity. To achieve this, virtual tags are created as follows: grids of a predefined size ($G$) are overlaid onto the region polygons, and the centers of these grids which are located within the region polygons are designated as the virtual tag locations. The sizes of the virtual tags are set to a small constant $\varepsilon$ to represent the absence of actual tags in these areas, as shown in Figure 3. For instance, we set $\varepsilon = 0.1\text{pt}$ in our approach. The grid size is determined while taking into account the sizes of subsequently placed tags. Specifically, the grid size ($G$) is defined as follows.

$$G = \frac{(F_{max} + F_{min})}{4} * \overline{L_{num}} \quad (4)$$

Here $F_{max}$ and $F_{min}$ represent the maximum and minimum font sizes of the tags, respectively, as specified by user settings in Yang et al. (2019)'s algorithm. $\overline{L_{num}} = \frac{\sum L_{num}}{N_{tag}}$, where $L_{num}$ denotes the number of letters in a tag, and $N_{tag}$ indicates the total number of tags. In subsequent iterative steps for placing tags, if a virtual tag falls within the area occupied by a placed tag, we delete the virtual tag from consideration.

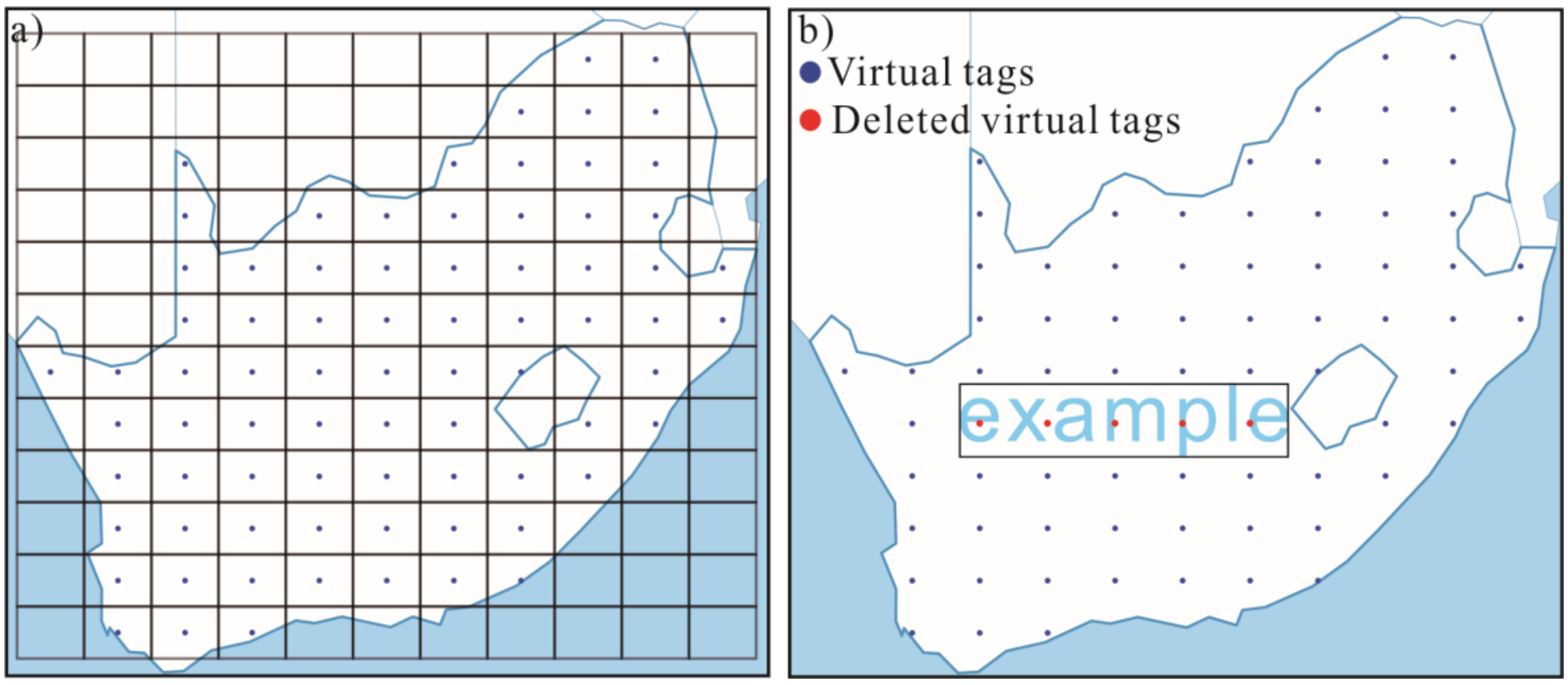

Figure 3. (a) Create virtual tags; (b) Delete the virtual tags in each iterative step if they occupy the area of a placed tag.

### 4.3 Filter out bad candidates

Our defined index aims to enhance the multi-scale visualization capability of tag maps. Before selecting the optimal location that best satisfies our defined index within the TIN-based tag layout algorithm, it may be necessary to filter out certain candidate locations, and filtering these candidates can also improve efficiency.

Initially, our approach builds upon the TIN-based method introduced by Yang et al. (2019) by iteratively examining triangular subareas to pinpoint suitable candidate locations for each tag. Given our objective of achieving uniform tag distances, we prioritize areas offering ample space for tag placement. Consequently, we focus our iteration process solely on the top $N_T$ triangular subareas with the largest areas, as they provide more space for tag placement. This selective approach also enhances efficiency compared to traversing all triangular subareas. Secondly, to maintain uniform distances between tags, we assess the distances between newly placed tags and those already positioned during each iteration. For instance, if a new tag were to be placed at a candidate location ($Loc_i$) and positioned too closely to previously placed tags compared to other candidate locations, we identify this proximity and eliminate $Loc_i$ from the current candidate location list. Therefore, we utilize two strategies to filter out undesirable candidates, as follows.

**Strategy I**. Instead of traversing all triangular subareas, we only iterate through the top $N_T$ triangular subareas based on their areas to identify candidate locations.

**Strategy II**. After identifying $M$ candidate locations denoted as $S_{\text{c-loc}}=\{Loc_i\}_{i=1}^{M}$ using the TIN-based algorithm in an iteration, we remove the locations that are situated too close to previously positioned tags. The term 'close' is determined relative to the average distance between tags (Wei et al., 2018), as follows: Suppose placing a current tag at location $Loc_i$ results in the shortest distance between previously placed tags, denoted as $D_i$. If $D_i < \bar{D}$, location $Loc_i$ is considered too close to previously positioned tags. $\bar{D} = \frac{\sum D_i}{M}$, and the distance here between two tags is calculated based on the distance between their centers for ease of calculation.

### 4.4 Select the best candidate

After identifying several suitable candidate locations, the selection of the optimal candidate location may hinge not only on its ability to support multi-scale visualization but also on other factors. For example, the candidate locations may place tags with different orientations, and users may have specific preferences for these orientations. Cognitive experiments according to Yang et al. (2021) have suggested that horizontally oriented tags may facilitate better comprehension of the tag map. Therefore, when tags can be placed with various orientations, it becomes necessary to weigh the preference for tag orientations alongside the capability for supporting multi-scale visualization during candidate selection. This preference for orientation can be considered by assigning a weight (Wei et al., 2018), as follows.

$$I' = \begin{cases} I * W_i & (I \geq 0) \\ I / W_i & (I < 0) \end{cases} \tag{5}$$

$W_i$ represents the weight assigned by the user to orientation $i$, $I$ is the defined index in Equation (3) of supporting multi-scale visualization. The default weights for each orientation are set to 1. Users can assign a larger $W_i$ to a specific orientation if they prefer that orientation. We offer nine fixed orientations including $[0°, 30°, 45°, 60°, 90°, -30°, -45°, -60°, -90°]$ for users to select, consistent with the approach outlined in Yang et al. (2019).

### 4.5 The algorithm to generate the initial tag layout

The iterative process for tag placement within region polygons is delineated in **Algorithm 1**. For further elucidation on the TIN-based tag map, comprehensive details can be found in the work of Yang et al. (2019).

---

**Algorithm 1. The iterative process for creating a tag layout**

---

**Input**:

- Region polygons;
- Sorted tags in descending order based on their values as $S_{tag} = \{Tag_1, Tag_2, ..., Tag_m, ...\}$;
- Maximum and minimum font sizes as $F_{max}$ and $F_{min}$;
- The threshold to filter out bad candidates and improve the efficiency as $N_T$;

**Output**: The tag layout $S'_{tag} = \{Tag'_1, Tag'_2, ..., Tag'_m, ...\}$;

1. Create the virtual tags;
2. **Do**

    a. Set the size of $Tag_1$ as $F$ according to the approach of Yang et al. (2019);

    b. Initialize $i \leftarrow 0$, $S_{c\text{-}loc} \leftarrow \varnothing$;

    c. Construct a TIN based on the region polygons and sort all the triangles in TIN in a descending area order as $S_{tri} = \{Tri_1, Tri_2, ..., Tri_m, ...\}$;

    d. **Do**

        **If** $Tag_1$ can be placed in $Tri_1$ **Then**

If $S_{\text{c-loc}}$ is ***Null*** **Then** $i \leftarrow 0$;

Add the center of $Tri_1$ as a location to $S_{\text{c-loc}}$;

Remove $Tri_1$ from $S_{\text{tri}}$ and update $S_{\text{tri}}$;

$i \leftarrow i+1$;

e. **While** ($i < N_T$ **AND** $S_{\text{tri}}$ is not ***Null***)

f. **If** $S_{\text{c-loc}}$ not ***Null*** **Then**

Filter out bad candidates in $S_{\text{c-loc}}$;

Select the best candidate from $S_{\text{c-loc}}$ ;

place $Tag_1$ at the selected location as $Tag'_1$, and add $Tag'_1$ to $S'_{\text{tag}}$;

g. Update the virtual tags;

h. Remove $Tag_1$ from $S_{\text{tag}}$ and update $S_{\text{tag}}$;

3. **While** ($F > F_{\min}$ AND $S_{\text{tag}}$ is not ***Null***)

4. **Return**: $S'_{\text{tag}} = \{Tag'_1, Tag'_2, ..., Tag'_m, ...\}$

### 4.6 Layout strategies under different map scales

Based on **Algorithm 1** presented in Section 4.5, we can generate the initial tag layout. To generate tag layouts across different scales, we follow the strategies delineated by Yang et al. (2019).

(1) Generate the initial tag layout

Given an initial set of tags and a geographical region, users are required to define their desired maximum and minimum font sizes as $F_{\max}$ and $F_{\min}$, respectively, and to set an initial scale as $S_{\text{initial}}$. Aligned with Yang et al. (2019)'s approach where font size is linearly correlated with a tag's weight value, the font size $F_i$ for a tag with a weight value $W_i$ is determined using Equation (6).

$$F_i = F_{\min} + (W_i - W_{\min}) \times (F_{\max} - F_{\min}) / (W_{\max} - W_{\min}) \quad (6)$$

Where $W_{\max}$ and $W_{\min}$ are the highest and lowest weight values among all tags, respectively. Upon calculating the font sizes for all tags, the tags are sequentially positioned within the geographical region according to their font sizes to create the initial tag layout at scale $S_{\text{initial}}$, as outlined in **Algorithm 1**. Furthermore, two things need to be noted here due to the limited space of the geographical region on the screen. Firstly, $F_{\max}$ may be excessively large, making it impossible to fit any tag within the region. In such cases, we iteratively decrease $F_{\max}$ by 1 ($F_{\max}= F_{\max}-1$) until the placement of the first tag is feasible. Secondly, as tags are placed within the region sequentially, the available space may diminish to the point where it becomes insufficient to accommodate all remaining tags. In this scenario, we continue to place tags until either no space remains or no further tags can be positioned.

(2) Generate the tag layouts across scales

It is essential to adjust the font sizes of the tags, as the screen representation of a geographical region also alters with changes in map scale. In accordance with the methodology proposed by Yang et al. (2019), we dynamically adjust each tag's font size from $F_i$ to $F'_i$ as the map scale changes, as dictated by Equation (7).

$$F'_i = F_i \times S_{\mathrm{tar}} / S_{\mathrm{ori}} \tag{7}$$

$S_{\mathrm{ori}}$ is the original map scale, and $S_{\mathrm{tar}}$ is the target scale when the user zooms out. Here the font sizes ($F_i$ and $F'_i$) determine the height of the lettering, which aligns with common practices in computer science. If $F'_i < F_{\min}$, the tag will be concealed from the screen. Conversely, the tag will reappear. If the font size of the largest tag is less than $F_{\min}$, the layout algorithm will not be initiated. In cases where the map is continuously zoomed in and the available whitespace within the region is sufficient to accommodate additional tags, a 'reconstruction' of the tag map is warranted. This is because some tags that are not included in the initial layout algorithm may now be displayed. It is noteworthy that the computational efficiency of using Equation (7) for font size adjustment is significantly higher than that of employing a reconstruction algorithm. Therefore, if all tags are initially placed using the layout algorithm, their font sizes can be exclusively adjusted using Equation (7), irrespective of whether the map is zoomed in or out. This enhances the speed at which the tag map can be displayed during zoom operations.

## 5. Experiment

### 5.1 Implementation details

(1) Datasets

To ensure comparability with previous research, we utilized the same dataset as Yang et al. (2019) for depicting topics relevant to countries worldwide. The foundational country map data was sourced from OpenStreetMap. Tags representing themes including ‘Culture’, ‘History’, and ‘Economy’ were extracted from Wikipedia data. Each word was treated as a tag, with its frequency of occurrence serving as the weight value. The initial scale for each region was determined according to its area and the screen resolution. Once a region was selected for tag map production, it was zoomed to full screen, with the scale being the ratio at which the region was displayed fullscreen.

(2) Experimental environment

We implemented our proposed approach within the Tag Map Explorer web-based application, which is built on OpenLayers and Turf.js and was initially developed by Yang et al. (2019). The map projection was set to Web Mercator Projection (EPSG:3857), which is the default projection in OpenLayers. The updated version of this application, integrating our approach, is also publicly accessible via our GitHub repository. The experiments were conducted on a personal computer equipped with an AMD Ryzen 7-7840HS Radeon 780M Graphics @3.80 GHz CPU,16GB RAM, and a screen resolution of 1920 × 1080.

(3) Evaluation metrics

As per Barth et al. (2014), six common metrics are typically employed to evaluate tag cloud layouts: realized adjacencies, distortion, compactness, uniform area utilization, aspect ratio, and running time. Of these metrics, realized adjacencies and distortion are mainly relevant to semantic tag map layouts, while uniform area utilization and aspect ratio are essentially the same for intrinsic tag maps. Consequently, we selected compactness (*C*) and running time (*t*) as evaluation metrics for this approach. Compactness denotes the extent of area utilized for displaying the words relative to the total area and is defined as $C = A_{\text{word}}/A_{\text{entire}}$, where $A_{\text{word}}$ is the sum of bounding box areas of all words, and $A_{\text{entire}}$ is the area of fitting regions. Additionally, since each tag occupies an area, a higher number of placed tags (*N*) indicates a more efficient utilization of map space. Therefore, we also used the count of placed tags (*N*) to evaluate compactness. Larger values of *C* or *N* indicate better compactness. Moreover, the defined negative spatial autocorrelation index (*I*) was adopted to gauge the capability of supporting multi-scale visualization. A larger *I* value indicates a better capability for supporting multi-scale visualization.

## 5.2 Results on benchmark datasets

### 5.2.1 Quantitative analysis

We provide two examples in two different kinds of region polygons with different settings of tag orientations (more examples can be created interactively by users via our GitHub repository). **Example 1**: The United States of America. In this example, tags are horizontally placed within a large region polygon characterized by an irregular shape, including islands. This demonstration validates the capability of creating tag layouts in polygons with complex shapes. The parameter settings used are as follows: $F_{\max}$=60pt, $F_{\min}$=6pt, $N_{\mathrm{T}}$=200, and an initial scale of 1:20,921,196, determined by the region's area and screen resolution. **Example 2**: Italy. In this example, tag orientations are adjusted based on the location of their corresponding triangles within a narrow region polygon. Given the narrowness of the polygon, checking all possible orientations can be time-consuming. The parameter settings used are as follows: $F_{\max}$=300pt, $F_{\min}$=6pt, $N_{\mathrm{T}}$=50, and an initial scale of 1: 3511885, determined by the region's area and screen resolution.

Furthermore, since Yang et al. (2020) indicated that tags with horizontal directions can enhance a better understanding of the tag map, we assigned a larger weight of 2 to the orientation in Equation (5).

As shown in Table 1, it is evident that implementing the strategy outlined in Section 4 leads to improvements in multi-scale visualization ability ($I$) in both the tag layouts of the USA and Italy, by 0.479 and 0.399, respectively, and with $I$>0. However, this enhancement may come at the expense of increased time or reduced tag placement. For instance, in the case of the USA layout, both approaches can accommodate 75 tags, but our method requires an additional 3.19 seconds. Conversely, for the Italy layout, our approach accommodates only 80 tags, whereas Yang et al. (2019)'s approach can accommodate 90 tags. However placing fewer tags can also reduce the time cost, resulting in a reduction in time by 9.33 seconds. Moreover, while maintaining the same number of tags in the USA layout, our approach achieves a higher compactness, increasing from 0.505 to 0.515. Conversely, in the Italy layout, our approach achieves a smaller compactness, decreasing from 0.616 to 0.538.

To further assess the efficiency of our proposed approach, we compared it with two similar methods by Buchin et al. (2016) and Reckziegel et al. (2018). Buchin et al. (2016)'s algorithm required 30 minutes and 60 minutes for the two different datasets. Reckziegel et al. (2018) reported that their aPTM models took at least 13.6 seconds, while their ePTM models took up to 7281.5 seconds. In contrast, the maximum time consumed by our approach is 27.69 seconds, and in certain cases, it runs in less than 10 seconds.

These findings indicate that our proposed strategy efficiently enhances multi-scale visualization capabilities to a certain extent but may entail additional time or fewer placed tags. Specifically, by maintaining the same number of tags, our approach achieves higher compactness, and placing fewer tags also leads to improved efficiency.

Table 1. Statistical results on two example areas.

| | ***Region*** | ***N*↑** | ***I*↑** | ***C*↑** | ***t*↓** |
|---|---|---|---|---|---|
| The USA | Approach of Yang et al. (2019) | 75 | -0.435 | 0.505 | 5.24s |
| | Our proposed approach | 75 | 0.044 | 0.515 | 8.43s |
| Italy | Approach of Yang et al. (2019) | 90 | -0.348 | 0.616 | 27.69s |
| | Our proposed approach | 80 | 0.051 | 0.538 | 9.33s |

**Note**. *N* represents the tag count in the tag layout, *I* denotes the defined index for evaluating the multi-scale visualization ability of the tag map, *C* indicates the compactness, and *t* indicates the running time of the algorithm. The direction of the arrow indicates whether the indicator's value is better when it is larger or smaller.

### 5.2.2 Qualitative analysis

The evaluation of multi-scale visualization ability can be challenging due to its subjective nature. To facilitate a qualitative analysis, we present the multi-scale visualization results for the USA and Italy in Figures 4 and 5 for our approach and Yang et al.'s (2019) approach. As illustrated in Figure 4, our approach demonstrates the capability to position small tags in narrow areas that are distant from previously placed tags. Conversely, Yang et al. (2019)'s method tends to favor areas with larger spaces if the region polygon is large enough, facilitating the placement of larger tags in proximity. This distinction is notably observed in Area A, where Yang et al. (2019)'s approach places few tags. This effect becomes more pronounced with decreasing scale, as evidenced by the emptying of Area B at level 2, with larger tags concentrating in the center of the polygon at level 3. In contrast, our proposed approach achieves a more even distribution of tags of varying sizes across the region polygon. At level 2, Area B retains tags, and at level 3, tags continue to be scattered throughout the polygon. Note that the northeastern area of the Lower 48 in Figure 4 is also empty at level 3 in our approach -- this occurs because the region is too narrow to accommodate large tags horizontally. Therefore, when users navigate to a very small scale (level 3), this area will inevitably appear empty after the removal of small tags in both our approach and Yang et al. (2019)'s approach. Similar observations are evident in Figure 5 for Italy. With decreasing scale, the tag layout generated by Yang et al. (2019) tends to concentrate in the center of the polygon at level 3, leaving empty areas such as Areas D and E. Conversely, our approach ensures a more even distribution of tags across the polygon at all levels. It should also be noted that our approach may leave small island polygons empty when users navigate to a smaller scale. For example, in Figure 5, Sicily appears empty at level 3 in our approach, whereas it is not in Yang et al. (2019)'s approach. This occurs because the sub-index for each polygon is computed first when calculating the overall index to evaluate multi-scale visualization capability, making small tags more likely to be placed in polygons with small areas. However, leaving these small island polygons empty may be visually acceptable, and user preferences regarding the emphasis on particular small areas could be considered in future work. These observations underscore the superior uniformity in tag size distribution achieved by our proposed approach compared to existing methods, thereby enhancing multi-scale visualization capabilities.

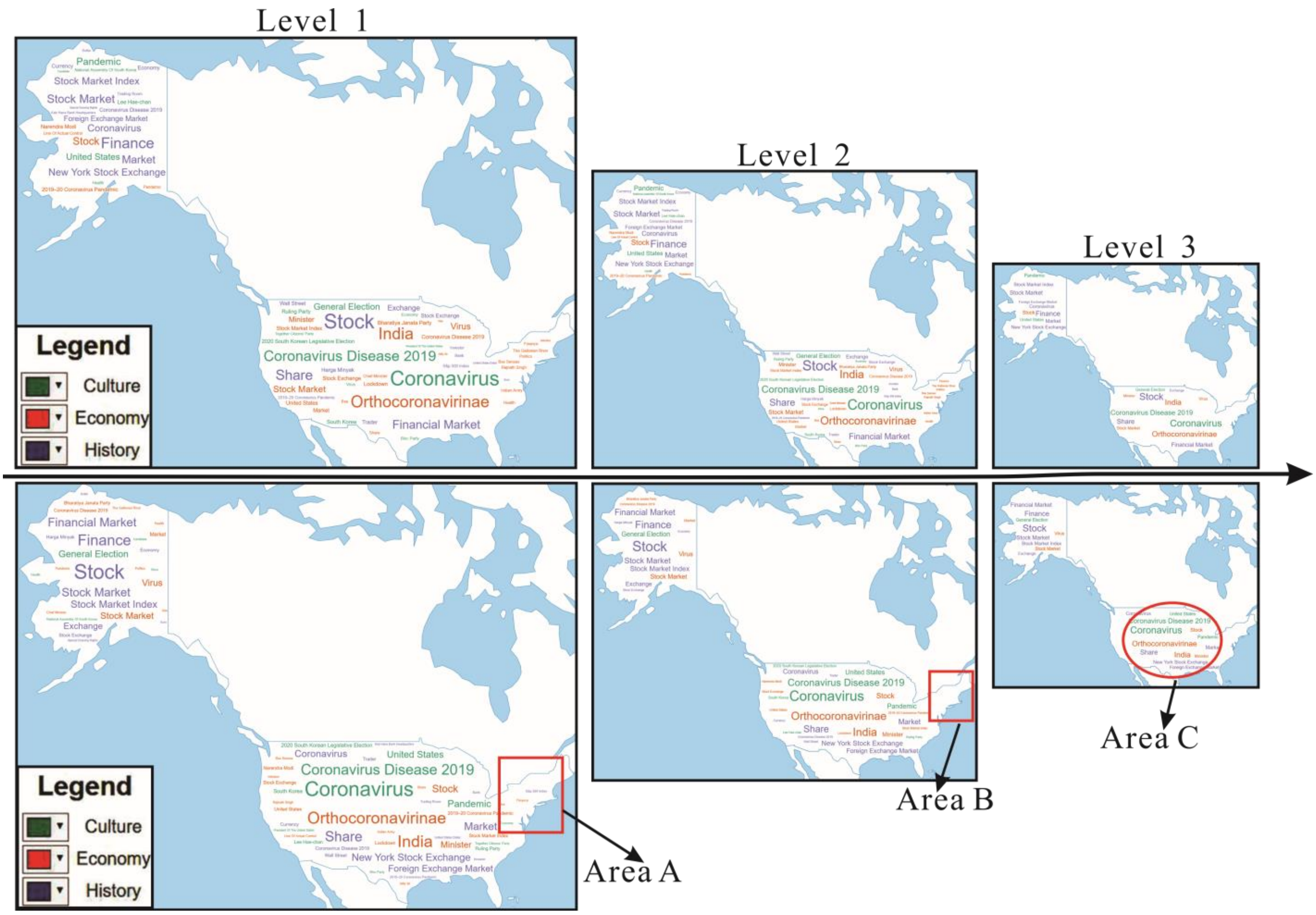

Figure 4. Multi-scale visualization results of tags for the USA. The top row displays results generated by our proposed approach, while the bottom row shows results generated by the approach of Yang et al. (2019).

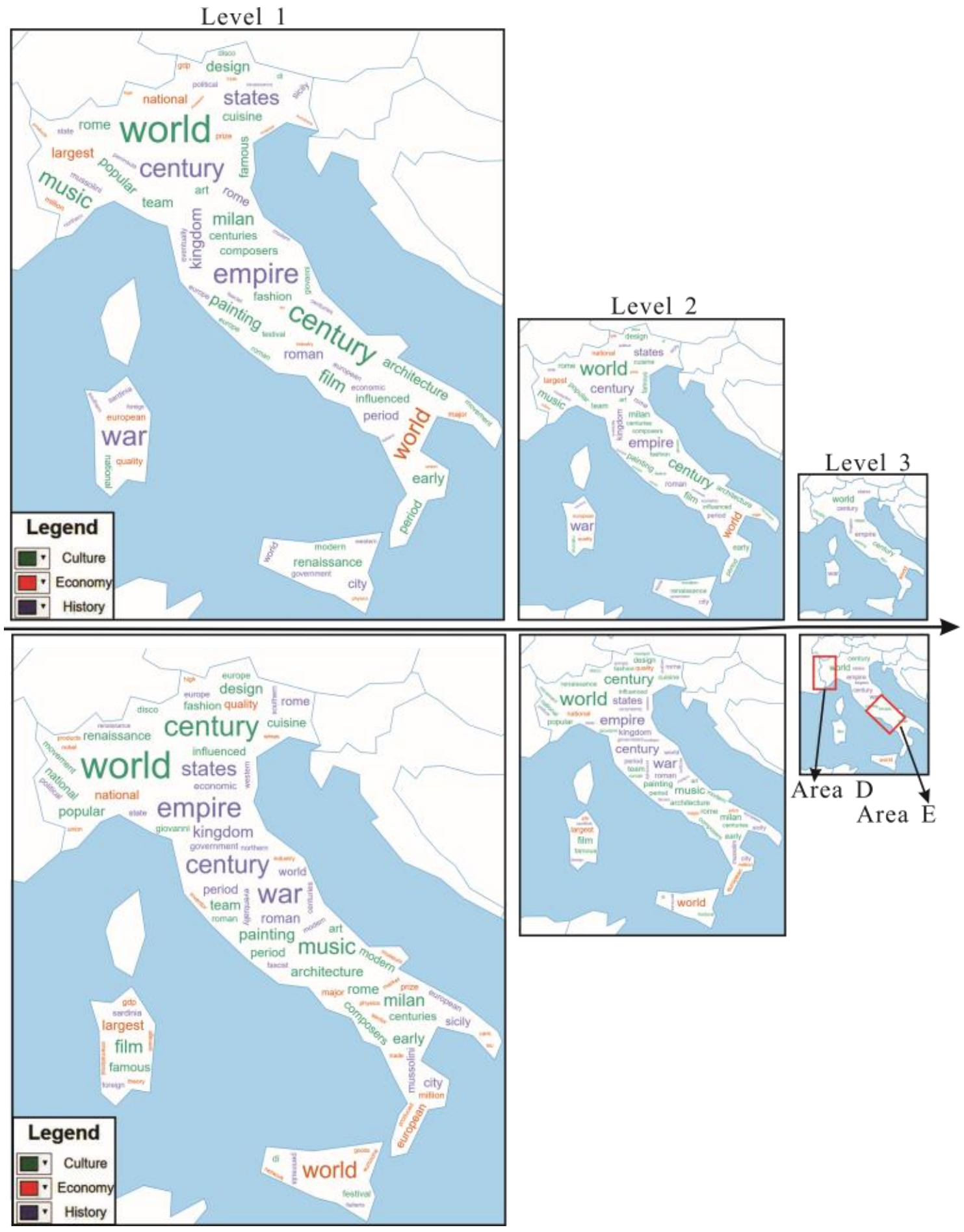

Figure 5. Multi-scale visualization results of tags for Italy. The top row displays results generated by our proposed approach, while the bottom row shows results generated by the approach of Yang et al. (2019).

To further facilitate qualitative analysis, we also provide the intermediate stages of placing the tags in Italy, as shown in Figure 6. From Figure 6, we can observe that large tags tend to gather in close proximity in Yang et al. (2019)'s approach, which is particularly noticeable when $N$=6 or $N$=14 in the upper part of the region. This clustering occurs because Yang et al. (2019)'s approach always aims to place tags in the center of the triangle with the largest area. In contrast, our approach seeks the best location for a tag among multiple large-area triangles by considering the capability for multi-scale visualization. As a result, the tags are more evenly distributed across the region polygon. When users explore the intrinsic map at a smaller scale, large tags are less likely to cluster together or create large empty areas after smaller tags are dynamically removed from the layout. Therefore, our approach provides better support for multi-scale visualization compared to Yang et al. (2019)'s approach.

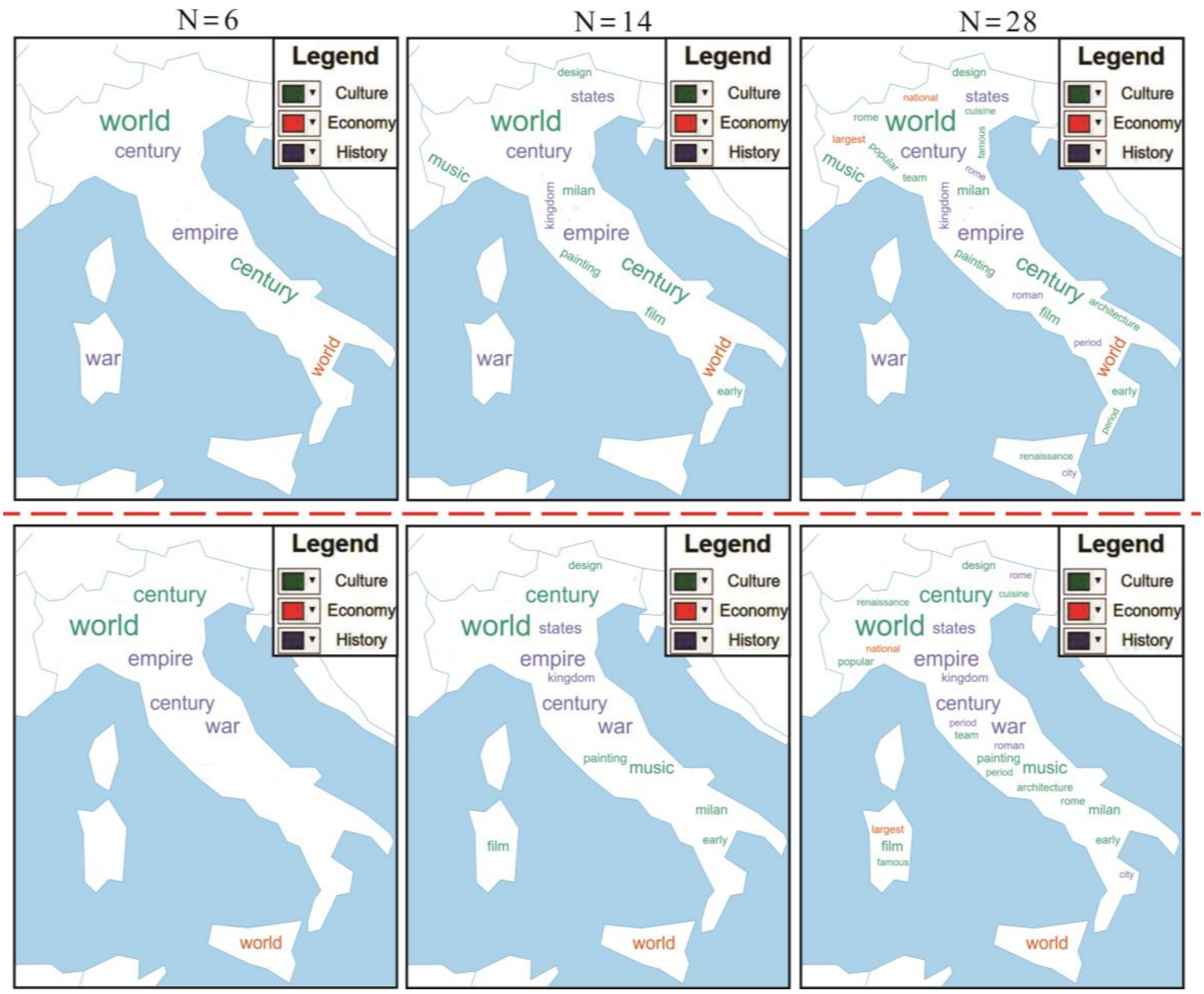

Figure 6. The intermediate stages of placing the tags in Italy. The top row displays results generated by our proposed approach, while the bottom row shows results generated by the approach of Yang et al. (2019).

### 5.3 Effectiveness analysis of the strategies for creating virtual tags

Virtual tags are introduced in our approach to compute the predefined spatial autocorrelation index. In ensuring homogeneity when no tags are initially placed within the region polygon, we adopt a grid-based sampling strategy to create virtual tags. However, alternative strategies can also be considered; for instance, randomly generating virtual tags positioned far from the region polygon can similarly ensure homogeneity (Liu et al., 2014). Thus, we applied both strategies to generate layouts for our two example areas (the USA and Italy) using identical parameter settings as in Section 5.2. The statistical results are presented in Table 2, and the visualization outcomes are depicted in Figure 7.

From Table 2, it is evident that both strategies exhibit distinct advantages in tag layout. For the tag layout of the USA, employing random-based sampling resulted in the placement of 82 tags with an index value of 0.940 and a compactness value of 0.554, achieved in 24.84 seconds. Conversely, grid-based sampling placed 75 tags, yielding a lower index value of 0.044 and a smaller compactness of 0.515, albeit taking less running time of 8.43 seconds to complete. The results for the tag layout of Italy exhibited minor variations. With random-based sampling, the algorithm placed 90 tags with an index value of 0.164 and a compactness value of 0.614, albeit with a runtime of 42.37 seconds. In contrast, grid-based sampling placed only 80 tags, resulting in a slightly higher index value of 0.051, but completed significantly faster and smaller compactness, requiring only 6.95 seconds and achieving 0.538 in compactness. However, two strategies can both improve the index of multi-scale visualization by comparing Tables 2 and 1. Comparing the two visualization results in Figure 7 with those in Figures 4 and 5 (top tows at level 1), it is apparent that the two sampling strategies both achieve a uniform distribution of tag sizes. These observations suggest that neither strategy exhibits clear superiority over the other. The random-based sampling strategy achieves a higher index value and greater compactness, but the grid-based sampling strategy requires less time. The substantial difference in index values between the two strategies is partly due to the scaling weight of 10 applied in Equation (3). The choice between random-based and grid-based sampling may hinge on practical considerations, such as the desired tag density and acceptable runtime constraints.

Table 2. Statistical results on two example areas by creating virtual tags via different strategies. The direction of the arrow indicates whether the indicator's value is better when it is larger or smaller.

| | ***Region*** | ***N*↑** | ***I*↑** | ***C*↑** | ***t*↓** |
|---|---|---|---|---|---|
| The USA | Random-based sampling | 82 | 0.940 | 0.554 | 24.84s |
| | Grid-based sampling | 75 | 0.044 | 0.515 | 8.43s |
| Italy | Random-based sampling | 90 | 0.164 | 0.614 | 42.37s |
| | Grid-based sampling | 80 | 0.051 | 0.538 | 9.33s |

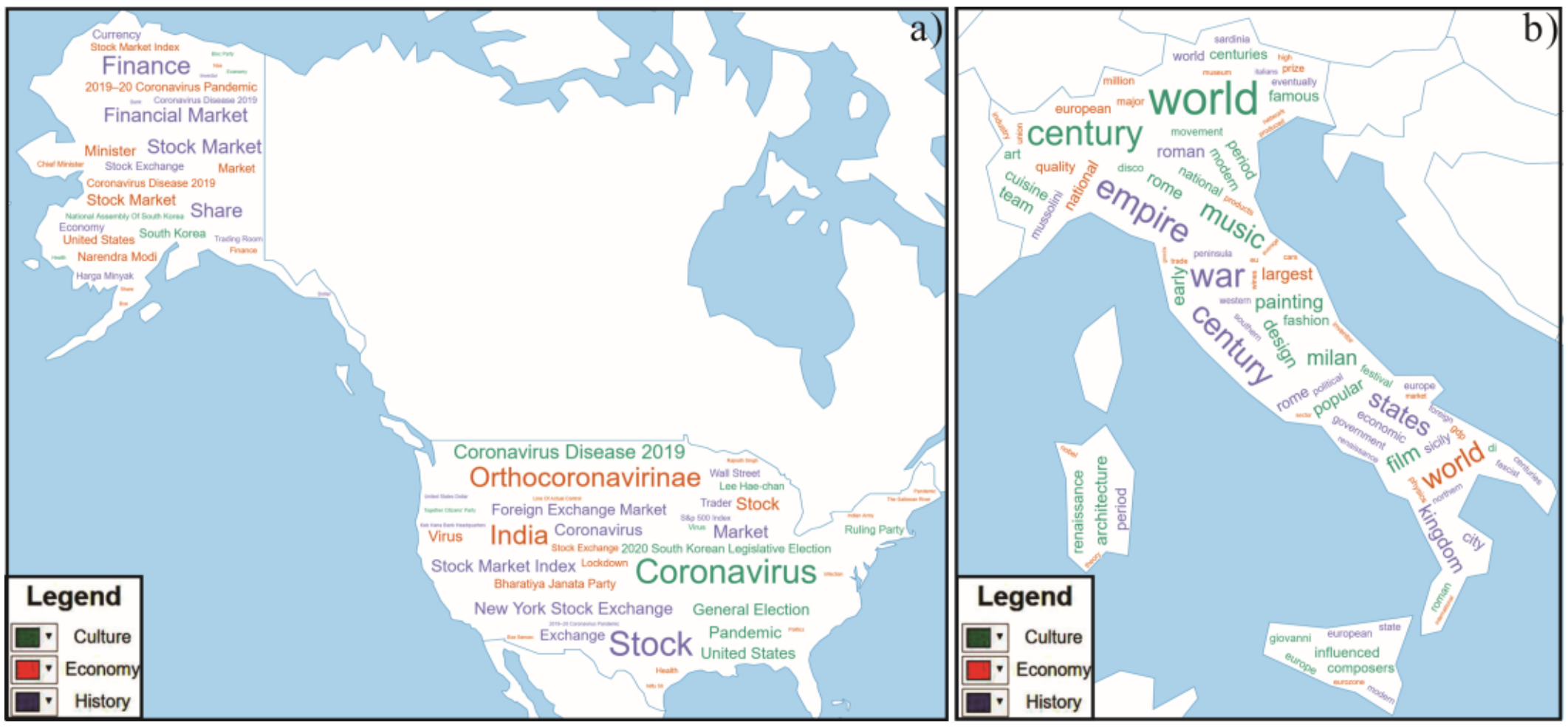

Figure 7. Visualization results of creating virtual tags using a random-based sampling strategy. (a) The tag map for the USA, compared with the grid-based sampling strategy for creating virtual tags as shown in Figure 4 (top row at level 1); (b) The tag map for Italy, compared with the grid-based sampling strategy for creating virtual tags as shown in Figure 5 (top row at level 1).

### 5.4 Effectiveness analysis of the strategies for filtering out bad candidates

Two strategies for filtering out unsuitable candidates are implemented in our approach to enhance tag layout quality and efficiency. **Strategy 1** involves iterating only through the top $N_T$ triangular subareas, while **Strategy 2** removes locations that are too close to previously positioned tags. To assess the effectiveness of these strategies, we utilized our proposed approach as a baseline and conducted ablation studies by individually removing each strategy. These studies were conducted on two example areas (the USA and Italy) with identical parameter settings as in Section 5.2 in which the tag densities are relatively high. As our proposed two strategies try to maintain the uniform tag distance, it will be more effective when tag density is low. Thus, we also conducted another ablation study in Egypt with a lower tag density. The parameter settings used for Egypt are set as follows: $F_{max}$=100pt, $F_{min}$=6pt, $N_T$=20, and an initial scale of 1: 2,670,772 determined by the region's area and screen resolution. The statistical results are presented in Table 3, while the visualization outcomes for Egypt are depicted in Figure 8.

As presented in Table 3, the absence of **Strategy 1** results in significantly longer processing times, being 4.04 times, 17.10 times, and 10.69 times longer than the baseline for the tag layouts of the USA, Italy, and Egypt, respectively. However, it also leads to an increase in the number of placed tags by 12, 10, and 0 and an increase in compactness by 0.016, 0.078, and 0, respectively. Moreover, as illustrated in Figure 8, omitting **Strategy 1** results in the generation of very large empty areas (Area A), particularly when tag density is low. This outcome stems from our approach's objective to optimize the defined index (*I*) and achieve the largest values of *I* as 0.707, 0.402, and 0.122, but potentially causing tags to concentrate in localized areas when **Strategy 1** is not applied. Balancing effectiveness and efficiency, these findings highlight **Strategy 1**'s crucial role in filtering out unfavorable candidates and facilitating a more optimized tag layout efficiently, even if it may reduce the *I* value.

Regarding **Strategy 2**, it yields comparable processing times to the baseline but with no evident advantages in metrics *N*, *I* and *C*. Furthermore, as depicted in Figure 8, the absence of **Strategy 2** results in the emergence of several empty areas (Area B and Area C), particularly under low tag density conditions. This occurrence arises from **Strategy 2**'s objective to eliminate candidate locations where the placement of a current tag would be too close to previously placed tags. Without **Strategy 2**, tags tend to prioritize achieving the best-defined index (*I*), potentially leading to their clustering around nearby tags. These results highlight the efficacy of **Strategy 2** in screening out undesirable candidates and facilitating the improvement of tag layout quality.

Table 3. Statistical results on the ablation studies. The direction of the arrow indicates whether the indicator's value is better when it is larger or smaller.

| | ***Region*** | ***N*↑** | ***I*↑** | ***C*↑** | ***t*↓** |
|---|---|---|---|---|---|
| | Baseline without strategy I | 87 | 0.707 | 0.531 | 34.04s |
| The USA | Baseline without strategy II | 53 | -0.406 | 0.453 | 2.37s |
| | Baseline | 75 | 0.044 | 0.515 | 8.43s |
| | Baseline without strategy I | 90 | 0.402 | 0.616 | 159.52s |
| Italy | Baseline without strategy II | 85 | 0.024 | 0.566 | 10.28s |
| | Baseline | 80 | 0.051 | 0.538 | 9.33s |
| | Baseline without strategy I | 90 | 0.122 | 0.093 | 45.01s |
| Algeria | Baseline without strategy II | 90 | -0.020 | 0.093 | 4.38s |
| | Baseline | 90 | 0.062 | 0.093 | 4.21s |

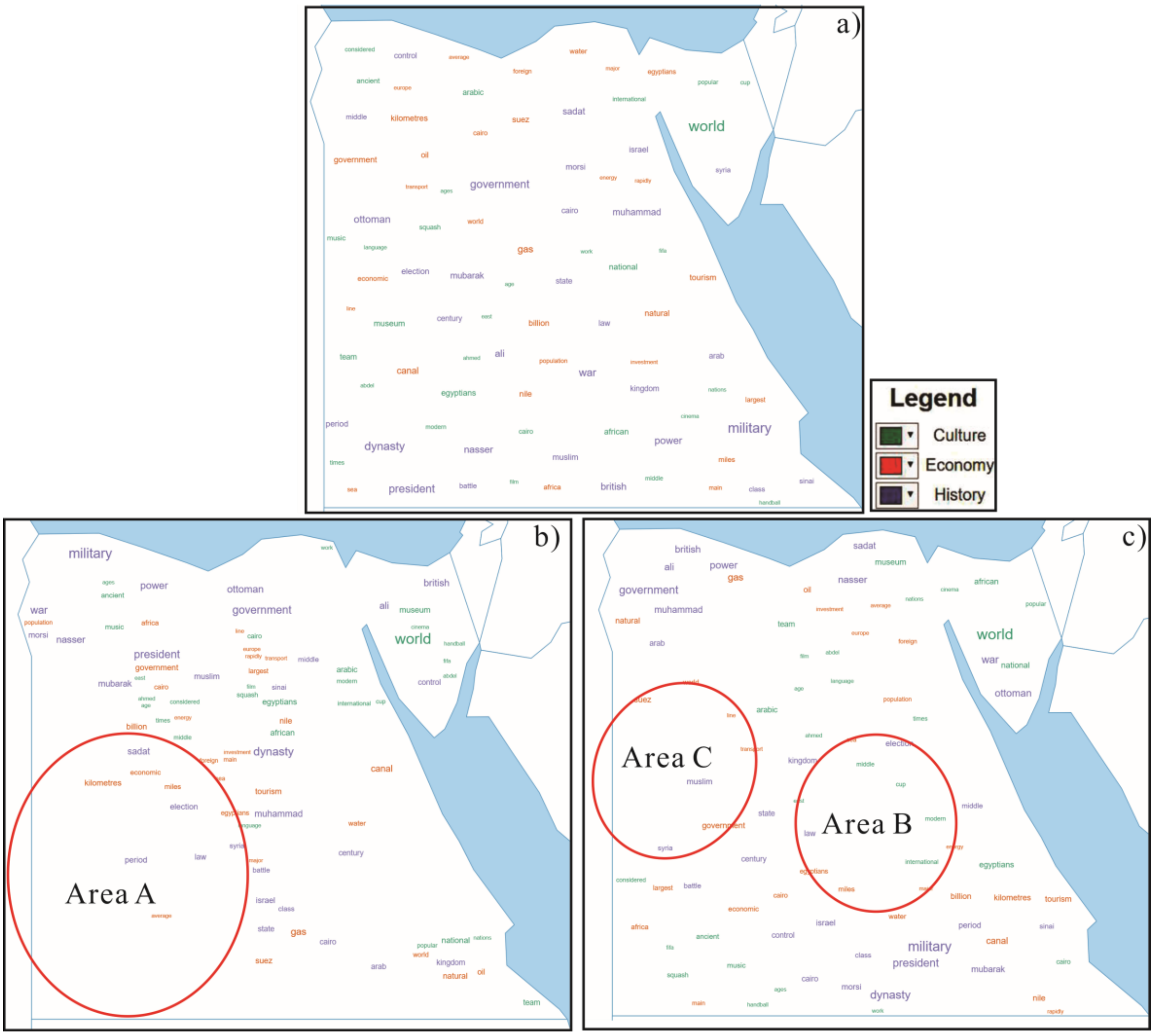

Figure 8. Visualization results of ablation results for Egypt. (a) Baseline; (b) without **Strategy 1**; (c) Without **Strategy 2**.

## 5.5 Effectiveness analysis of the strategies for selecting the best candidate

To incorporate the preference for different tag orientations, we assign varying weights in Equation (5) to select candidate locations. In our experiments, tags with horizontal directions are assigned a higher weight, denoted as $W_{hori}$=2, while the weights for other tag orientations are set to 1. To evaluate this strategy, we established our proposed approach as a baseline and conducted ablation studies by uniformly weighting all tag orientations. Additionally, we compared our approach with that of Yang et al. (2019), wherein tags with horizontal direction are prioritized. These investigations were carried out in the Italy area, with tags set to different orientations, using parameters specified in Section 5.2. The results are summarized in Table 4 and Figure 9, where the number of tags with horizontal orientation is denoted as $N_{hori}$.

As depicted in Table 4 and Figure 9, Yang et al. (2019)'s approach yielded the highest number of tags with horizontal direction and achieved the maximum compactness of 0.616, as they consistently prioritize tags with this orientation. However, their approach exhibited the lowest values in metric *I* (-0.348) and consumed the most time (27.69 seconds). In contrast, setting $W_{hori}$=2 resulted in 3 additional tags and 15 more tags with horizontal orientations, leading to an increase of compactness of 0.011. Nonetheless, this also increased the processing time from 6.95 seconds to 9.33 seconds and decreased the metric I from 0.213 to 0.051. These findings indicate that assigning higher weights to preferred orientations can enhance the number of tags with preferred orientations and compactness. However, this enhancement comes at the cost of increased processing time and reduced multi-scale visualization capability.

Table 4. Statistical results on the results of Italy area by assigning weights to preferred orientation. The direction of the arrow indicates whether the indicator's value is better when it is larger or smaller.

| ***Region*** | | ***N*↑** | ***I*↑** | ***C*↑** | ***t*↓** | **$N_{hori}$↑** |
|---|---|---|---|---|---|---|
| Italy | Approach of Yang et al. (2019) | 90 | -0.348 | 0.616 | 27.69s | 67 |
| | Our proposed approach with $W_{hori}$=2 | 80 | 0.051 | 0.538 | 9.33s | 44 |
| | Our proposed approach without $W_{hori}$=2 | 77 | 0.213 | 0.527 | 6.95s | 29 |

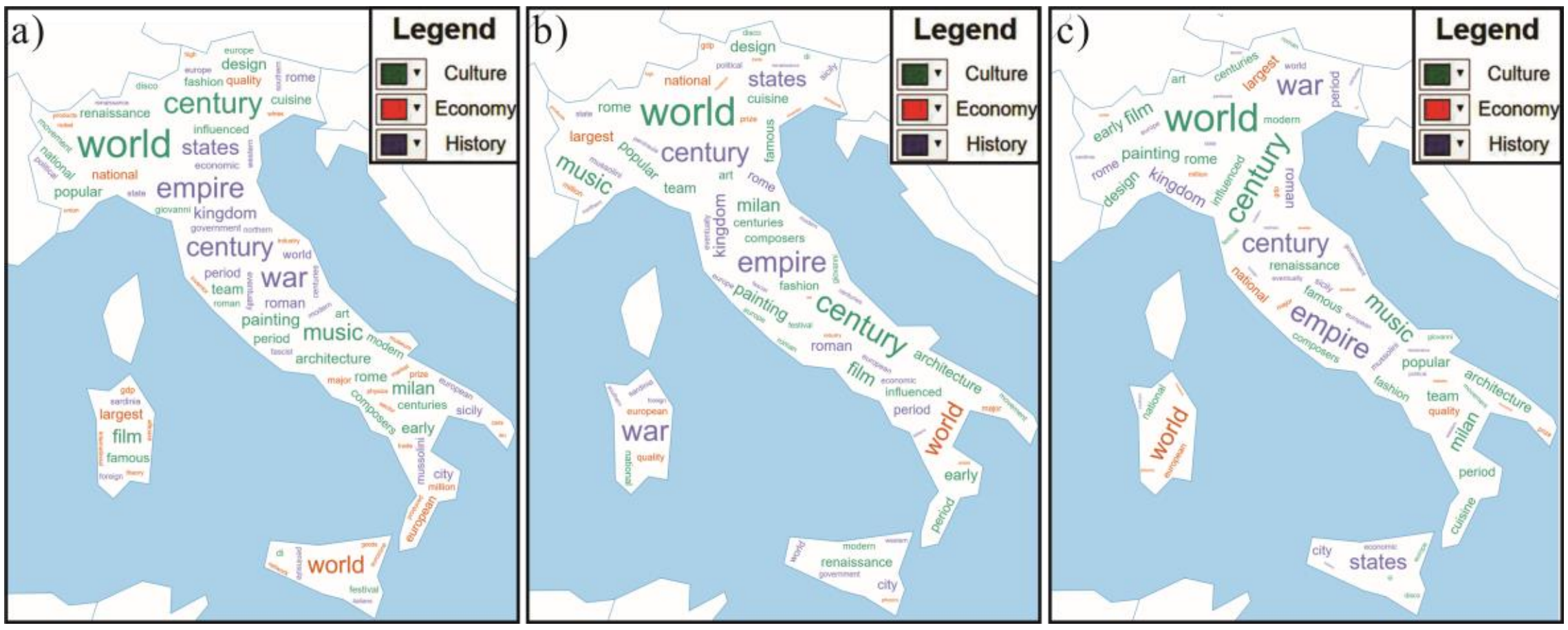

Figure 9. Visualization results of the tag map for Italy with different priorities of tag orientations. (a) Produced by using Yang et al. (2019)'s approach; (b) Produced by our proposed approach with $W_{hori}$=2; (c) Produced by our proposed approach with uniformly weighting all tag orientations.

## 6. Discussion

Spatial data exploration often requires users to navigate across multiple scales, necessitating support for multi-scale visualization (Zhang et al., 2018). In this study, we used the intrinsic tag map as an example and integrated the negative spatial autocorrelation index into the maps to evaluate and enhance their multi-scale visualization capability. Our results show that better visualization performance can be achieved when multi-scale visualization support is considered, despite trade-offs in compactness and time efficiency. These findings confirm the necessity of incorporating multi-scale visualization capability in spatial data visualization. However, visualization is also a user-driven task, and users should select settings based on their specific requirements, balancing efficiency and effectiveness. Our approach can also be extended to fulfil specific requirements (Section 6.1), and several limitations need to be addressed in future studies to improve this approach (Section 6.2).

### 6.1 *Extensions for specific user requirements*

Sometimes, users may have specific requirements, such as placing certain tags in specific positions or setting certain tags with specific font sizes. Our approach can also be extended to accommodate these special demands.

(1) Placing tags in specific positions. This can be achieved by initially placing the tags with predefined positions, ensuring they are set in their designated spots. Subsequent tags are then placed iteratively using our proposed approach. For example, a user may want to place the most important tag in Figure 4, ‘Coronavirus,’ in the center of the Lower 48 states of the USA. To fulfill this requirement, we can first place the tag ‘Coronavirus’ in the desired position and then iteratively place the remaining tags. The result is shown in Figure 10(a), illustrating that our approach can successfully produce a tag map with certain tags in specific positions.

(2) Setting tags with specific font sizes. This can be accomplished by assigning specific font sizes to the designated tags. The font sizes for other tags are determined using Equation (6). All tags are then placed iteratively according to their font sizes. For instance, a user may want to emphasize ‘Stock’ in Figure 4 and set tags containing ‘Stock’ with larger font sizes. To fulfill this requirement, we can determine the font sizes for all tags, setting the font sizes for tags containing ‘Stock’ to $F_{\text{user}}$. Where $F_{\text{user}}$ is the user-specified font size. Then all tags are placed iteratively in order of their font sizes using our approach. The result is shown in Figure 10(b), illustrating that our approach successfully produces a tag map with specific tags in specific font sizes. However, it should be noted that due to the limited space of the geographical region on the screen, $F_{\text{user}}$ may sometimes be excessively large, making it impossible to fit a relevant tag within the region. In such cases, we iteratively decrease $F_{\text{user}}$ by 1 until the placement of the tag is feasible according to Yang et al. (2019).

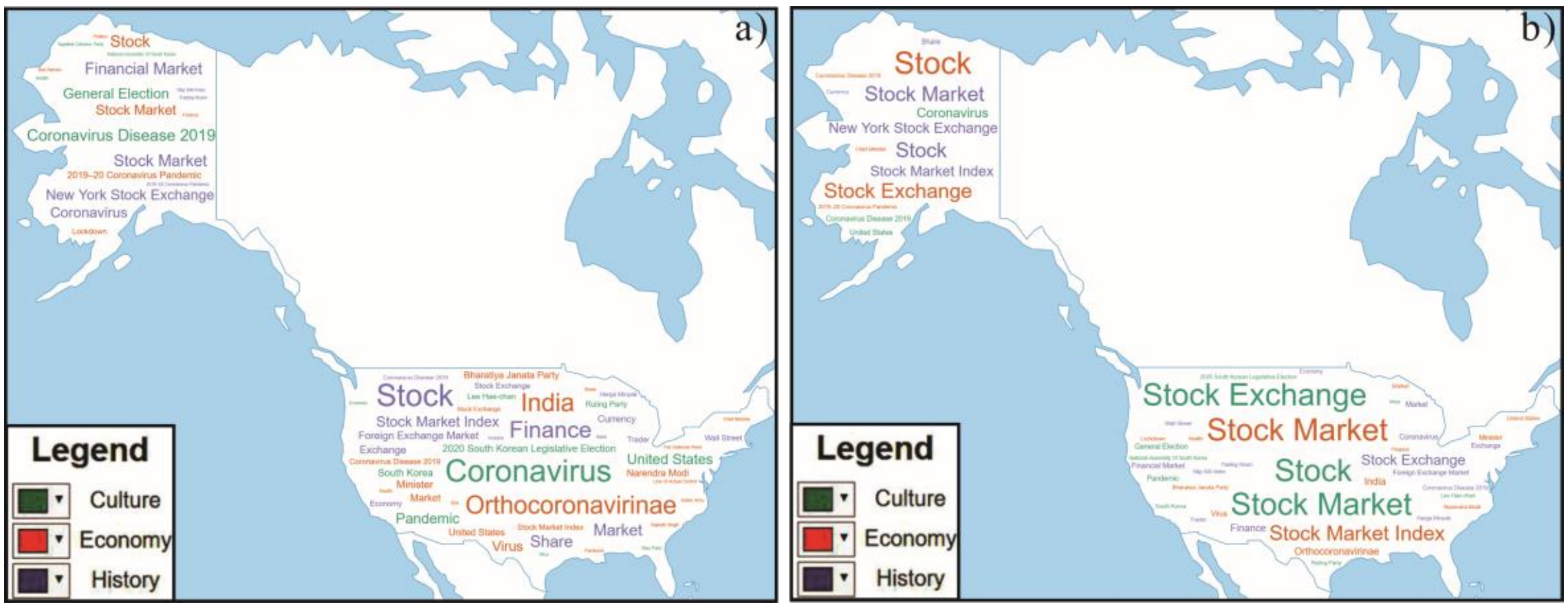

Figure 10. Visualization results of the tag map for the USA with different user demands: (a) Placing the tag ‘Coronavirus’ in the center of the Lower 48 states; (b) Setting tags containing ‘Stock’ with larger font sizes.

## 6.2 *Limitations*

(1) Multi-scale visualization capability is crucial for tag maps, but it is not the sole determinant of quality. Factors such as compactness and distribution are also significant. While our approach enhances multi-scale visualization, it may inadvertently compromise other factors. Thus, strategies that balance these considerations or cater to diverse user requirements may be necessary.

(2) We aim to enhance multi-scale ability by adjusting the initial tag layout, which remains fixed in subsequent user interactions. However, numerous other strategies supporting multi-scale visualization with user interaction may also be suitable, such as dynamically adjusting the layout or tag size based on user interactions or adopting alternative visualization methods. To develop a more flexible multi-scale visualization approach for tag maps, it is essential to integrate user demands, interactions, and alternative visualization forms rather than solely focusing on adjusting the initial tag layout.

(3) Parameter settings may vary depending on the shape of the fitting boundary and the distribution of tag data. Developing adaptive parameter settings tailored to different scenarios to achieve satisfactory results is imperative for future research.

## 7. Conclusion

To evaluate and improve the multi-scale visualization capability within the intrinsic tag map, we attribute this capability to the even distribution of tags with varying sizes across the region. We integrate the negative spatial auto-correlation index into tag maps to assess the uniformity of tag size distribution and, thus, evaluate the multi-scale visualization ability. Furthermore, we incorporate this index into a TIN-based tag map layout approach to enhance its ability to support multi-scale visualization by iteratively filtering out candidate tags and selecting optimal tags that meet the defined index criteria. Experimental results indicate that our proposed approach can indeed enhance multi-scale visualization capabilities when compared to existing methods, although trade-offs in compactness and time efficiency were observed. Specifically, when retaining the same number of tags in the layout, our approach achieves higher compactness but requires more time. Conversely, when reducing the number of tags in the layout, our approach exhibits reduced time requirements but lower compactness. Additionally, we discussed the effectiveness of various applied strategies aligned with existing approaches to generate diverse tag maps tailored to user preferences. Future work will focus on three key aspects: (1) Incorporating post-strategies, such as developing algorithms to dynamically displace tags to refine the tag map layout based on scale or exploring alternative visualization forms, such as providing multi-views of layouts at different scales, instead of solely adjusting the initial layout; (2) Enhancing algorithm efficiency by applying strategies such as applying spatial indexing to ensure scalability with larger datasets; (3) Developing user-centered algorithms that accommodate specific user demands, such as emphasizing particular small areas.

**Disclosure statement**

No potential conflict of interest was reported by the author(s).

**Data and code availability statement and data deposition**

The data and code that support the findings of this study are all openly available on our GitHub. The website is https://github.com/TrentonWei/Multi-scale-TagMap.git.

**Acknowledgments**